\documentclass[10pt,letterpaper,twocolumn,english,prx,superscriptaddress,showkeys,reprint]{revtex4-1}
\usepackage[T1]{fontenc}
\usepackage[latin9]{inputenc}
\usepackage{float}
\usepackage{amsmath}
\usepackage{amssymb}
\usepackage{cancel}
\usepackage{graphicx}

\makeatletter

\usepackage{hyperref}

\renewcommand\[{\begin{equation}}
\renewcommand\]{\end{equation}}

\makeatother

\usepackage{babel}
\begin{document}

\title{Signatures of a $4\pi$-periodic supercurrent in the voltage response
of capacitively shunted topological Josephson junctions}

\author{Jordi Picó-Cortés}

\affiliation{Instituto de Ciencia de Materiales de Madrid (CSIC), Spain.}

\author{Fernando Domínguez}

\affiliation{University of Würzburg, Institute for Theoretical Physics and Astrophysics
Am Hubland, Deutschland.}

\author{Gloria Platero}

\affiliation{Instituto de Ciencia de Materiales de Madrid (CSIC), Spain.}
\begin{abstract}
We investigate theoretical aspects of the detection of Majorana bound
states in Josephson junctions using a semiclassical model of the junction.
The influence of a $4\pi$-periodic supercurrent contribution can
be detected through its effect on the width of the Shapiro steps and
the Fourier spectrum of the voltage. We explain how the inclusion
of a capacitance term results in a strong quenching of the odd steps
when the junction is underdamped, which may be used to effectively
detect Majorana bound states. Furthermore, in presence of capacitance
the first and third steps are quenched to a different degree, as observed
experimentally. We examine the emission spectrum of phase-locked solutions,
showing that the presence of period-doubling may complicate the measurement
of the $4\pi$-periodic contribution from the Fourier spectrum. Finally,
we study the voltage response in the quasiperiodic regime and indicate
how the Fourier spectra and the first-return maps in this regime reflect
the change of periodicity in the supercurrent. 
\end{abstract}



\maketitle

\section{Introduction}

Topological degenerated states in quantum systems are subject of present
active research, both for their fascinating fundamental physical properties
and for the possibility of using them as a platform for topological
quantum computation~\citep{Tewari2007,Nayak2008,Bonderson2011,Jiang2011}.
Topological phases of superconductors which support Majorana bound
states(MBS)~\citep{Kitaev2001,Fu2008} can be implemented in solid
state setups presenting spin-orbit coupling, broken time-reversal
symmetry and superconductivity. Different experimental configurations
to detect MBS have been proposed~\citep{Bolech2007,Benjamin2010,Akhmerov2011,Beenakker2013}.
In particular, MBS can be detected in Josephson junctions~\citep{Tanaka2009,Pikulin2012,Tkachov2013},
through either zero-bias conductance peaks~\citep{Mourik2012} or
its effect on the current-phase relation in the dc Josephson effect~\citep{Olund2012}.
Recently, the current-phase relation in a Josephson junction formed
by one-dimensional nanowires featuring MBS has been observed experimentally
through the vanishing of the odd Shapiro steps in ac biased Josephson
junctions~\citep{Rokhinson2012,Wiedenmann2016,Bocquillon2016,1603.09611},
showing that this setup can be used to effectively detect MBS.

The appearance of Shapiro steps is one example of non-linear phenomena
in mesoscopic systems ~\citep{Okuyama1981,He1985,Pedersen1993,Ambika1997}.
Non-linear transport in different solid state systems has been analyzed
in the past~\citep{InstabSolidState,Jalabert1994,Bulashenko1995,Bonilla2005},
showing interesting regimes, such as quasiperiodicity\citep{Bohr1984,Jensen1984},
frequency locking\citep{0034-4885-59-8-001,10.2307/73893} and different
routes to chaos~\citep{He1984,Gwinn1987,Luo1998b,Alhassid2000}.
In that direction, one promising area of research focuses on the relationship
between topology and non-linearity. For example, the interplay between
topology and instabilities has been recently analyzed in bosonic systems
under ac driving~\citep{Engelhardt2016} and junction arrays mimicking
the SSH model~\citep{1611.01467}. The Shapiro experiment in a topological
Josephson junction has been theoretically analyzed by means of a semiclassical
RSJ model~\citep{Domnguez2012,1609.00372,1701.07389}, and with a
finite capacitance in the high ac-bias limit~\citep{Maiti2015}.
\begin{figure}[H]
\includegraphics[width=1\columnwidth]{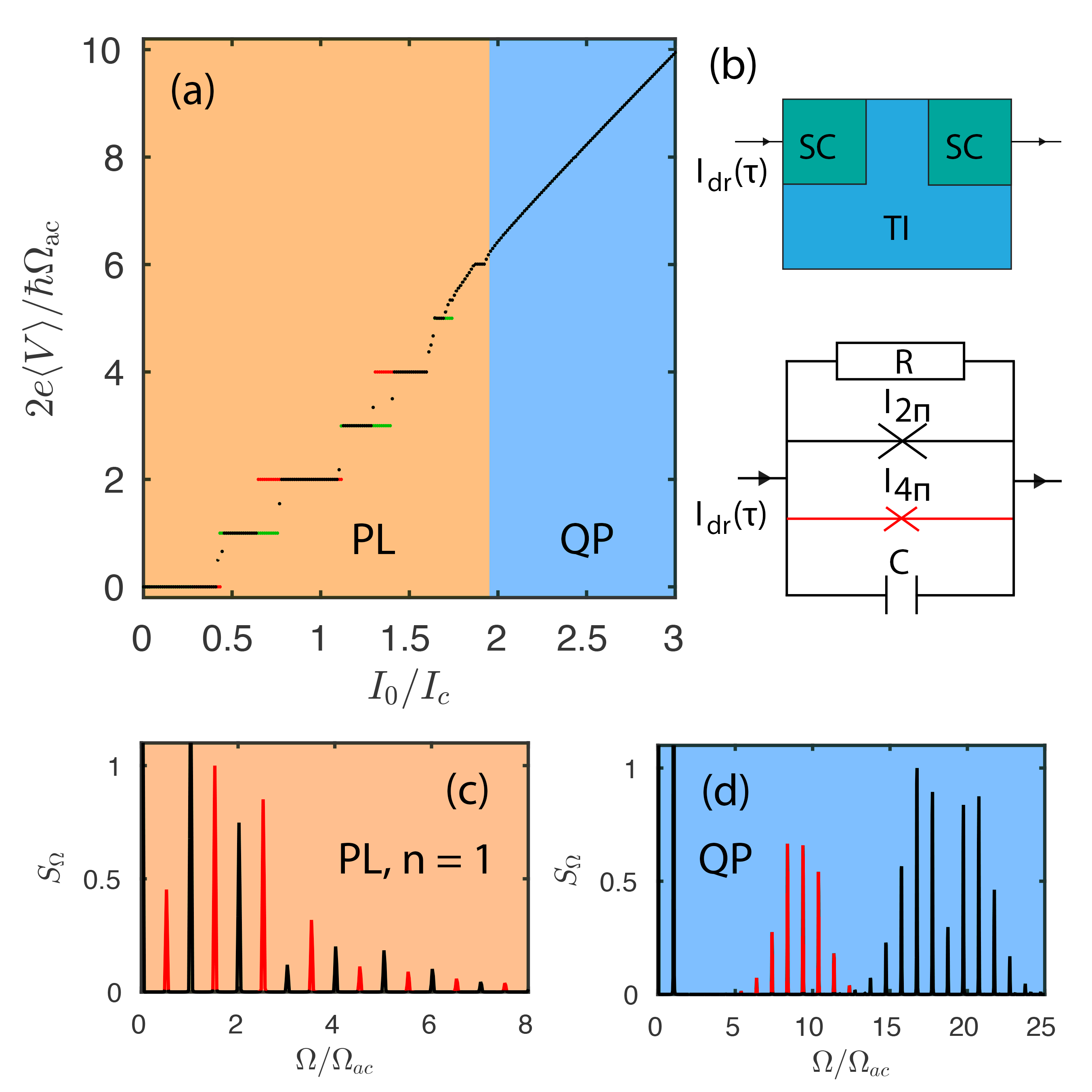}

\caption{(a) $I_{0}-\mathcal{h}V\mathcal{i}$ curve showing Shapiro steps corresponding
to the phase-locked regime (PL, orange) and a linear $I_{0}-\mathcal{h}V\mathcal{i}$
section for the quasiperiodic regime (QP, blue). The reduction (increase)
in the odd (even) steps for $I_{4\pi}=0.3I_{2\pi}$ is marked in green
(red). (b) Top: scheme of a topological Josephson junction. Superconductivity
is induced into a topological insulator (TI, blue) through the proximity
effect by a superconductor (SC, green). Bottom: circuit representation
of the junction in the RCSJ model, with elements corresponding to
the resistive (R), capacitive (C) and the ordinary ($I_{2\pi}$) and
$4\pi$-periodic ($I_{4\pi}$) supercurrent channels. (c), (d) Fourier
spectra from a PL solution in the first Shapiro step and from a QP
solution, respectively. The peaks in red are a result of the $4\pi$-periodic
supercurrent.\label{fig:(A)-Scheme-of-1}}
\end{figure}

In this work we investigate both the phase-locked and the quasiperiodic
regimes of a capacitively shunted Josephson Junction driven by an
ac current in presence of a $4\pi$-periodic supercurrent contribution.
This paper is organized as follows. In Sec.~\ref{sec:Theoretical-Model.}
we introduce the RCSJ model to describe such a system. In Sec.~\ref{sec:Shapiro-Experiments}
we study the influence of a $4\pi$-periodic contribution on the width
of the Shapiro steps and indicate the parameter regions where the
junction is strongly affected by the change in the periodicity of
the supercurrent. In Sec.~\ref{sec:Emission-Spectrum-Analysis} we
consider the possibility of measuring the emission spectrum of the
junction from both the phase-locked and quasiperiodic regimes in order
to detect MBS.

\section{Theoretical Model\label{sec:Theoretical-Model.}}

The study of the current-driven Josephson junction is a difficult
task from the microscopic point of view. It does not only involve
out of equilibrium processes but also strong Coulomb interactions
and dissipation. The problem becomes drastically simplified in the
semiclassical limit, yielding the resistively capacitively shunted
junction (RCSJ) model~\citep{McCumber1968}, represented schematically
in Fig.~\ref{fig:(A)-Scheme-of-1}~(b). This model describes the
evolution of the superconducting phase difference $\varphi$ by means
of the equation of motion $CdV/dt+V/R+I_{\mathrm{sc}}(\varphi)=I_{\mathrm{dr}}(\tau)$,
which results from equating the external current bias, which we take
as $I_{\mathrm{dr}}(\tau)=I_{0}+I_{1}\mathrm{sin}(\Omega_{\mathrm{ac}}\tau)$,
to a circuit consisting of three parallel channels: capacitive ($C$),
resistive ($R$) and supercurrent $I_{\mbox{sc}}(\varphi)$ channels.
We can eliminate $V$ in favor of $\varphi$ by using the Josephson
equation $V(\tau)=(2e/\hbar)d\varphi/d\tau$, yielding

\begin{equation}
\frac{d^{2}\varphi}{dt^{2}}+\sigma\frac{d\varphi}{dt}+i_{\mathrm{sc}}(\varphi)=i_{0}+i_{1}\mathrm{sin}(\omega_{\mathrm{ac}}t),\label{eq:rcsj eq}
\end{equation}
where $i_{k}\equiv I_{k}/I_{c}$, $k=0,1,\mathrm{sc}$ and $I_{c}=\mathrm{max}[I_{\mathrm{sc}}(\varphi)]$
is the critical value of the supercurrent. This expression has been
written in dimensionless units by defining a dimensionless time $t=\omega_{c}\tau$
and referring the ac bias frequency $\omega_{\mathrm{ac}}=\Omega_{\mathrm{ac}}/\omega_{c}$
in units of the plasma frequency $\omega_{c}\equiv\sqrt{2eI_{c}/\hbar C}$.
We have also introduced the damping parameter $\sigma\equiv\sqrt{\hbar/2eI_{c}R^{2}C}$
which gives the relative importance of the capacitive and resistive
channels. For $\sigma\gtrsim1$, the system is overdamped and the
effect of capacitance is negligible. For $\sigma\lesssim1$, the junction
is underdamped and the capacitance cannot be neglected.

\begin{figure}[t]
\begin{centering}
\includegraphics[width=1\columnwidth]{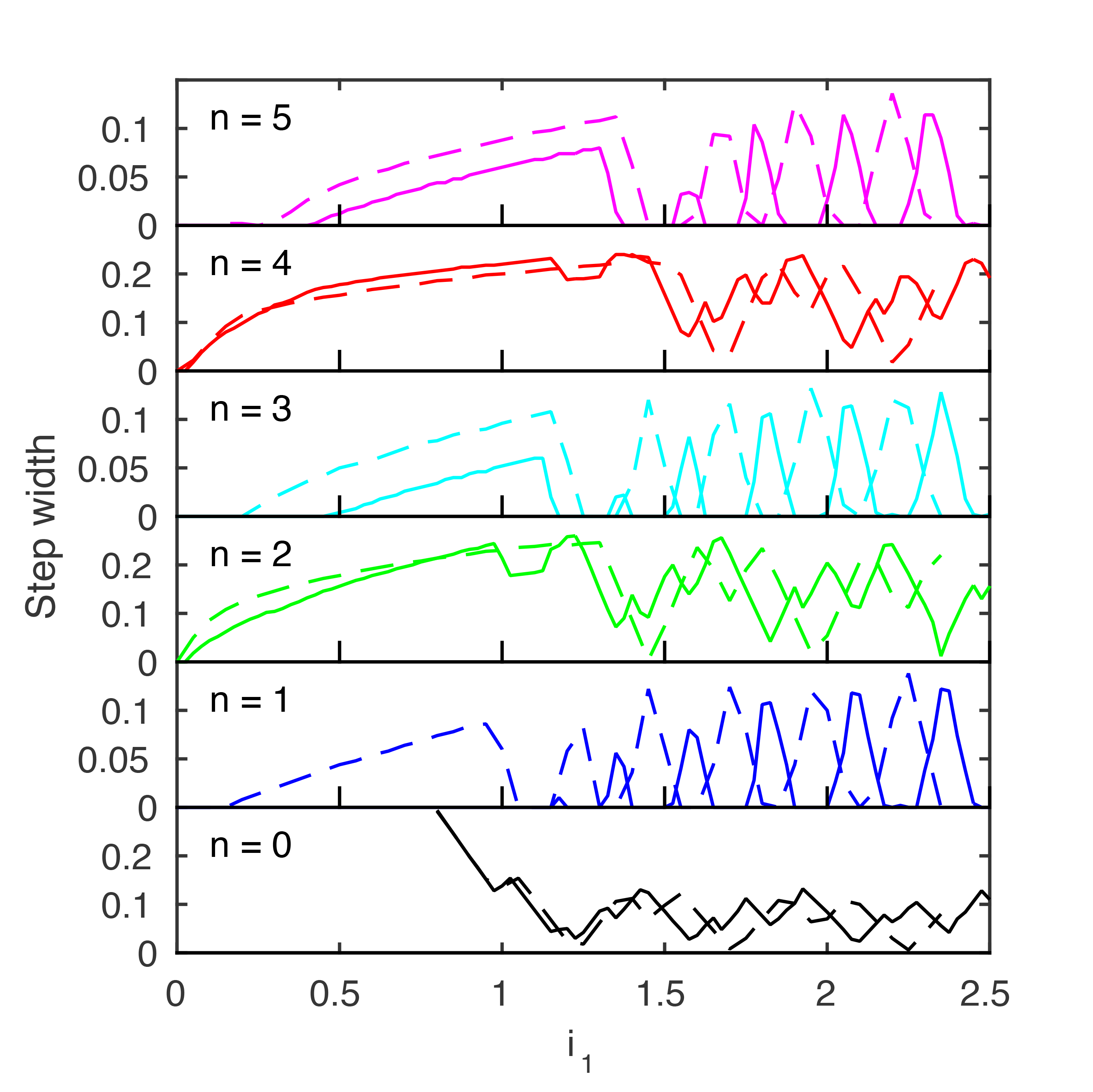}
\par\end{centering}

\caption{Calculated Shapiro step amplitudes as a function of $i_{1}$ for $\sigma\omega_{\mathrm{ac}}=0.1$
and $x=0.2$. The step amplitudes are normalized to the amplitude
of the zeroth step at $i_{1}=0$. The dashed curves correspond to
the RSJ model ($C=0$) and the solid curves to the RCSJ model with
$\sigma=1$. \label{figurei1}}
\end{figure}

In the presence of MBS, the supercurrent can be roughly described
by the sum of two contributions, $i_{\mbox{sc}}(\varphi)=i_{2\pi}\mathrm{sin}(\varphi)+i_{4\pi}\mathrm{sin}(\varphi/2)$,
where the first term corresponds to the usual $2\pi$-periodic supercurrent
and the second term is a $4\pi$-periodic contribution ($4\pi\mbox{SC}$),
which arises in the presence of MBS. Henceforth, we will characterize
the junction by the ratio $x\equiv i_{4\pi}/i_{2\pi}$. Note that
by writing $i_{2\pi}$ and $i_{4\pi}$ as constant coefficients we
neglect finite size effects, and all possible transitions towards
the quasicontinuum. 

The solution to Eq.~\ref{eq:rcsj eq} yields the induced voltage
$v(t)\equiv d\varphi/dt=(I_{c}R\sigma)^{-1}V$. In absence of an ac
bias, i.e: $i_{1}=0$, the voltage is a periodic function with frequency
$\omega_{0}=\mathcal{h}v(t)\mathcal{i}$, where $\mathcal{h}\ldots\mathcal{i}$
denotes time averaging. For $i_{1}\neq0$, the voltage is in general
a quasiperiodic function of frequencies $\omega_{0}$ and $\omega_{\mathrm{ac}}$.
When $\omega_{0}$ and $\omega_{\mathrm{ac}}$ are commensurate, the
system is said to be in phase-lock, and the average voltage is a multiple
of the ac bias frequency, i.e: $\mathcal{h}v(t)\mathcal{i}=n\omega_{\mathrm{ac}}$,
$n=0,1,2,\ldots$ . In Fig.~\ref{fig:(A)-Scheme-of-1}~(a) we have
represented the average voltage $\mathcal{h}v(t)\mathcal{i}$ as a
function of $i_{0}$ for for $\sigma=1$, $\omega_{\mathrm{ac}}=0.3$
and $i_{1}=0.75$. For a finite value of the ac bias amplitude $i_{1}$,
the induced voltage develops plateaus called Shapiro steps, at integer
multiples of $\omega_{\mathrm{ac}}$. Inside these plateaus, the voltage
is phase-locked to the ac bias. Shapiro steps can be used to discriminate
the presence of MBS, because in the case of a pure $4\pi$-periodic
supercurrent, one would expect to observe only the Shapiro steps for
$n$ even. We will show below how a finite capacitance can give rise
to a more involved Shapiro step picture where odd steps may appear
at $i_{2\pi}=0$. Quasiperiodic solutions correspond roughly to the
linear sections~\footnote{Inside the linear sections of the $i_{0}-\mathcal{h}v\mathcal{i}$
curve, quasiperiodic solutions are interlocked with phase-locked ones.
Because irrational numbers appear infinitesimally close to rational
numbers, it is difficult to make a clear statement about the nature
of a particular solution in these regions. } of the curve at high $i_{0}$. Alternatively, the periodicity can
be studied directly from the Fourier spectrum of the signal. For the
phase-locked regime, the spectrum is changed according to the step,
as noted in Appendix~\ref{sec:Subharmonic-response-in}. For the
quasiperiodic regime, the presence of a $4\pi\mbox{SC}$ induces new
Fourier components at $\omega_{0}/2$. As an example, we show in Fig.~\ref{fig:(A)-Scheme-of-1}~(c)
and (d) the Fourier spectra for the two regimes.

\section{Shapiro Step Widths\label{sec:Shapiro-Experiments}}

Loosely speaking, the width of the $n$th Shapiro step as a function
of the ac bias $i_{1}$ follows the shape of the $n$th Bessel function,
as noted. However, in the presence of both $2\pi$ and $4\pi$-periodic
contributions to the supercurrent, this profile is qualitatively modified~\citep{Domnguez2012}.
For the RSJ model, ($C=0$), in the low ac bias amplitude regime ($i_{1}\ll1$),
the odd steps are suppressed provided that $i_{1}\lesssim i_{4\pi}$
and $\sigma\omega_{\mathrm{ac}}=\hbar\Omega_{\mathrm{ac}}/2eI_{c}R\lesssim i_{4\pi}$.
Furthermore, in the high ac bias amplitude regime ($i_{1}\gg1$) the
even steps show a beating pattern coming from the contribution of
both supercurrent terms~\citep{1701.07389}. This high ac bias amplitude
behavior persists for intermediate values of the capacitance. This
is the case of Fig.~\ref{figurei1}, where we show the Shapiro step
width as a function of the ac bias amplitude $i_{1}$ for $\sigma=1$. 

The presence of a finite capacitance modifies the Shapiro step widths
drastically as the junction is taken into the underdamped regime,
$\sigma\lesssim1$, as noted in the Appendices~\ref{sec:Phase-lock-in-the}
and~\ref{sec:Outside-the-Bessel-1}. In this regime, the range of
frequencies where we observe only even steps has to be reconsidered.
In addition to $\sigma\omega_{\mathrm{ac}}=\hbar\Omega_{\mathrm{ac}}/2eI_{c}R\lesssim i_{4\pi}$
we also require that $\omega_{\mathrm{ac}}^{2}=\hbar\Omega_{\mathrm{ac}}^{2}C/2eI_{c}\lesssim i_{4\pi}$.
These conditions have been derived in the Appendix~\ref{sec:Outside-the-Bessel}
from the effect of a $4\pi\mbox{SC}$ on the voltage output. Remarkably,
the presence of capacitance extends considerably the condition $i_{1}\lesssim i_{4\pi}$,
valid for the RSJ model. In Fig.~\ref{figurei1} we have represented
the width of the first five Shapiro steps for $x=0.2$ and $\omega_{\mathrm{ac}}=0.1$
for both the RCSJ model with $\sigma=1$ (solid curves) and the RSJ
model (dashed curves). For $\sigma=1$, the first step vanishes for
ac bias amplitudes up to $i_{1}\simeq1.2$ much larger than the amplitude
of the $4\pi\mbox{SC}$, $i_{4\pi}=0.175$. Hence, the underdamped
junction is a useful platform for detecting MBS even when the $4\pi\mbox{SC}$
is a small fraction of the total supercurrent. In contrast to the
RSJ model, the quenching of the odd steps depends on the step number:
the third and fifth steps vanish only up to $i_{1}\simeq0.5$. This
occurs because, at higher voltages, the resistive term, which is proportional
to the voltage, is of greater importance than the capacitive one.
Hence, for higher steps the results for $\sigma=1$ and $\sigma\to\infty$
become more similar. These new conditions obtained from the RCSJ model
have to be considered when estimating $i_{4\pi}$ from the disappearance
of the odd Shapiro steps in experiments. 

\begin{figure}
\includegraphics[width=1\columnwidth]{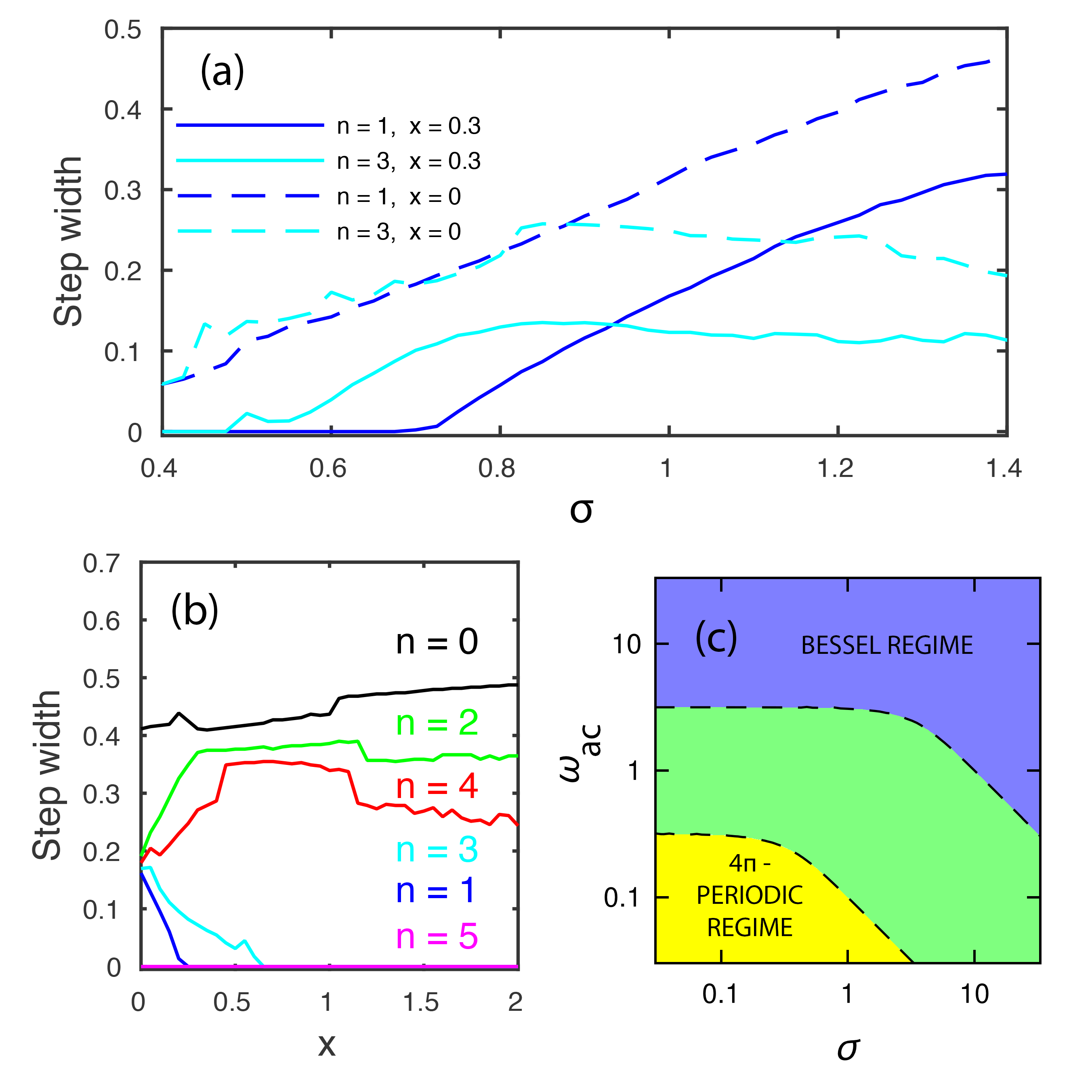}

\caption{(a) Odd step widths, $n=1$ (blue) and $n=3$ (cyan) as a function
of the damping parameter $\sigma$ for $\omega_{\mathrm{ac}}=0.3$,
$i_{1}=0.75$. (b) Step widths as a function of $x=i_{4\pi}/i_{2\pi}$
for $\sigma=0.65$. (c) Schematic phase diagram for the parameters
$\sigma$ and $\omega_{\mathrm{ac}}$ in logarithmic scale. The dashed
lines correspond to $\tilde{i}_{4\pi}=0.1$ and $\tilde{i}_{4\pi}=10$
(see text below) for $i_{4\pi}=1$. The yellow area corresponds to
the $4\pi$-periodic regime. The blue area corresponds to the Bessel
regime. The green area corresponds to the intermediate regime.\label{figureCAP}}

\end{figure}

In order to have an intuition about how the capacitance modifies the
odd step widths, we show in Fig.~\ref{figureCAP}~(a), the step
width of the first and third steps as a function of the damping parameter
$\sigma$ for a value of the ac bias $i_{1}=0.75$ larger than the
amplitude of the $4\pi\mbox{SC}$, $i_{4\pi}=0.175$. We see how decreasing
$\sigma$ results in the suppression of the first step, while the
third step vanishes for a smaller value of $\sigma$. In Fig.~\ref{figureCAP}~(b)
we have represented the step width as a function of the ratio $x=i_{4\pi}/i_{2\pi}$.
The first step vanishes for $x\simeq0.25$ while the third step requires
$x\simeq0.65$ to be suppressed.

On the other hand, when these conditions are not satisfied, the presence
of a finite $4\pi\mbox{SC}$ is not enough to suppress the odd steps.
In Appendix~\ref{sub:The-high-ac} we obtain that for high ac bias
amplitudes, such that
\[
\tilde{i}_{1}\gg\omega_{\mathrm{ac}}^{-4},\,\,\,\,\,\tilde{i}_{1}\gg\omega_{\mathrm{ac}}^{-2}\sigma^{-2},
\]
where $\tilde{i}_{1}=i_{1}(\omega_{\mathrm{ac}}\sqrt{\omega_{\mathrm{ac}}^{2}+\sigma^{2}})^{-1}$,
the junction is weakly affected by the change in the periodicity of
the supercurrent. Even for low ac bias amplitudes, if $\sigma\omega_{\mathrm{ac}}\gg1$
or $\omega_{\mathrm{ac}}^{2}\gg1$ most of the current will flow through
the resistive and capacitive channels and the effect of the $4\pi\mbox{SC}$
on the odd steps will be minimal. If any of these four conditions
is met, the junction is said to be in the Bessel regime. In this regime,
the Shapiro step widths can be obtained analytically, as noted in
the Appendix~\ref{sec:Phase-lock-in-the}.

In Fig.~\ref{figureCAP}~(c) we have represented an schematic phase
diagram for the RCSJ model with a $4\pi\mbox{SC}$ as a function of
the parameters $\sigma$ and $\omega_{\mathrm{ac}}$. The yellow region
corresponds to the regime where we expect the $4\pi\mbox{SC}$ to
have a strong effect on the junction behavior, the $4\pi$-\emph{periodic
regime}. The blue region corresponds to the Bessel regime and the
green region corresponds to the intermediate phase, where the odd
steps are suppressed only for low ac bias amplitude. The approximate
phase boundaries are determined by $\tilde{i}_{4\pi}=0.1$ and $\tilde{i}_{4\pi}=10$,
where we have defined $\tilde{i}_{4\pi}=i_{4\pi}\left[\omega_{\mathrm{ac}}\sqrt{\omega_{\mathrm{ac}}^{2}+\sigma^{2}}\right]^{-1}$.

\begin{figure}
\includegraphics[width=0.85\columnwidth]{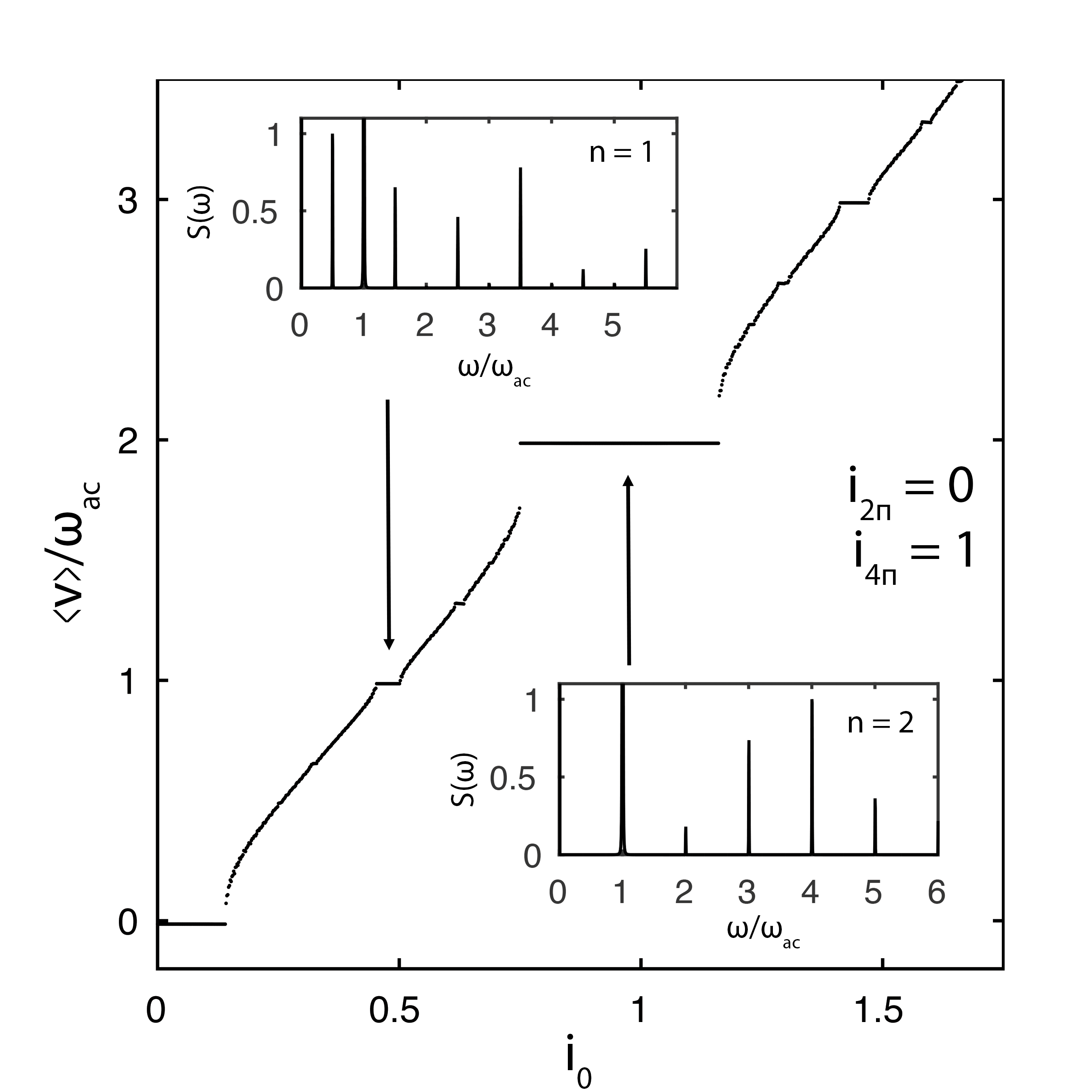}

\caption{Shapiro step curve for $i_{2\pi}=0$ and $i_{4\pi}=1$ ($x\to\infty)$.
Insets: Fourier spectrum obtained numerically at the first and second
steps. The Fourier spectrum in the first step consists only of half-harmonic
components plus a component at $\omega_{\mathrm{ac}}$, while the
Fourier spectrum at the second step consists only of components at
integer multiples of $\omega_{\mathrm{ac}}$, in accordance with the
analytical results of Appendix~\ref{sub:The-high-ac}. Parameters
considered: $\omega_{\mathrm{ac}}=1.6$, $\sigma=0.3$, $\tilde{i}_{1}=10$.\label{fig:remnant}}
\end{figure}

Another consequence of the presence of a finite capacitance is that
the odd steps do not necessarily disappear even if $i_{2\pi}=0$,
as illustrated by the results of Fig.~\ref{fig:remnant}. Odd steps
may still appear as a consequence of subharmonic phase-locking. Subharmonic
steps, such as in Fig.~\ref{fig:remnant} are always of smaller width
than the corresponding harmonic steps~\citep{10.2307/73893}. This
type of behavior is known to happen in the RCSJ equation as a consequence
of symmetry breaking in the non-linear supercurrent term. This possibility
is shown explicitly to happen in the high ac-bias regime in Appendix~\ref{sub:The-high-ac}.
In the RSJ limit, i.e: $C=0$, subharmonic phase-lock is rigorously
forbidden~\citep{Waldram1982}. Moreover, in the presence of strong
step overlap, numerical results indicate that subharmonic steps are
strongly quenched. Since step overlap occurs predominantly when $\sigma\omega_{\mathrm{ac}}\ll\tilde{i}_{1}^{2}$~\citep{0034-4885-59-8-001},
subharmonic steps seldom appear in the $4\pi$-periodic regime as
defined above. 

\section{Emission Spectrum Analysis\label{sec:Emission-Spectrum-Analysis}}

The periodicity of the response can be studied through the frequency
spectra, $S_{\omega}=|v(\omega)|$, where $v(\omega)$ is the Fourier
transform of the signal obtained in a Shapiro experiment. The emission
spectrum of the voltage was obtained in Ref.~\citep{1603.09611}
from experiments on a topological junction, in order to probe the
phase dependent periodicity of the junction as a function of the dc-current
bias, $i_{0}$. Below, we will analyze in detail the Fourier spectrum
of the voltage in the presence of both dc and ac bias: $i_{\mathrm{dr}}(t)=i_{0}+i_{1}\mathrm{sin}(\omega_{\mathrm{ac}}t)$.
We consider the Fourier spectrum of both phase-locked and quasiperiodic
solutions.

\begin{figure}
\includegraphics[width=1\columnwidth]{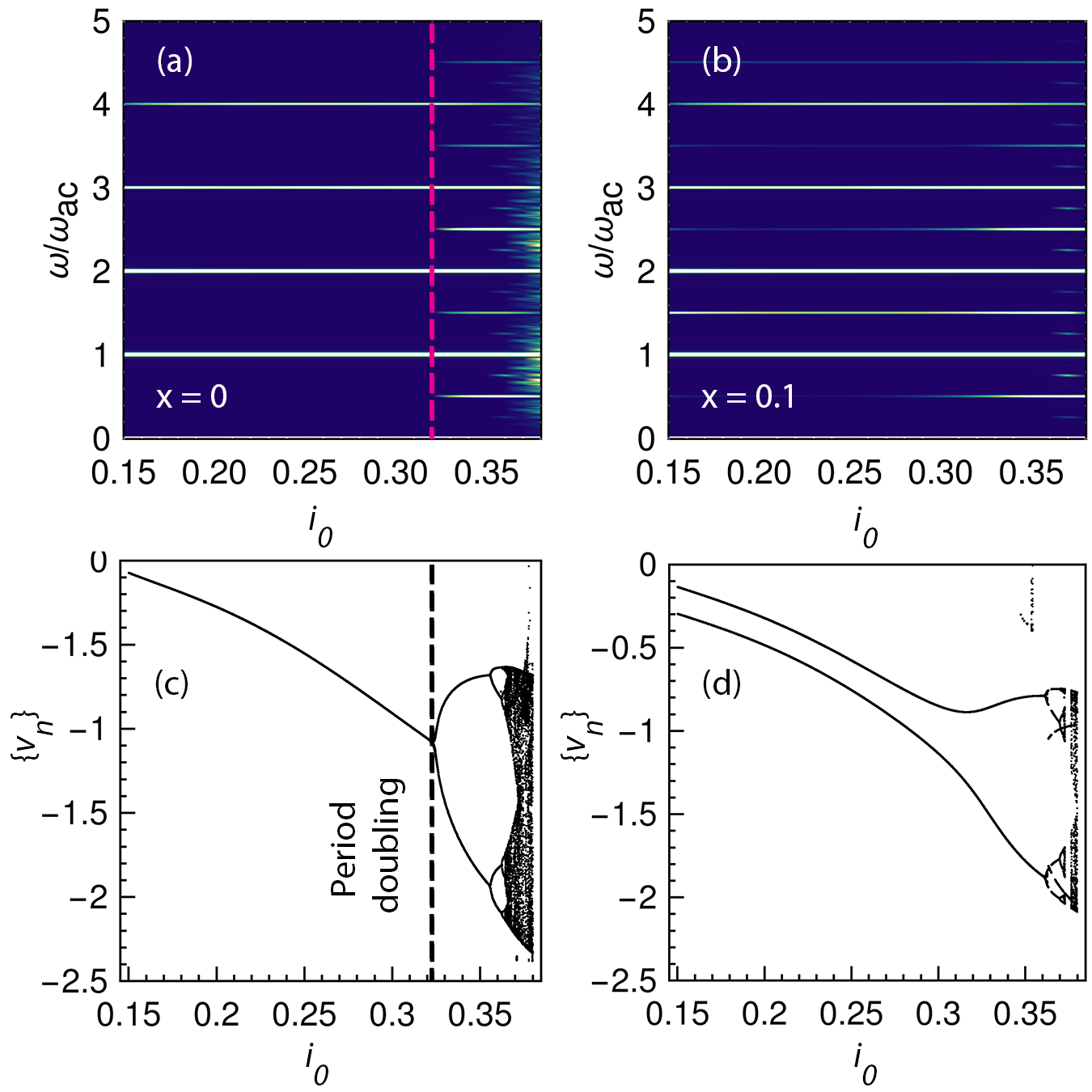}

\caption{(a) Density plot of the Fourier transform of the voltage signal $v(\omega)$
for $\sigma=0.3$, $\omega_{\mathrm{ac}}=0.6$, $x=0$ and (b) $x=0.1$.
(c) Bifurcation diagram for $\sigma=0.3$, $\omega_{\mathrm{ac}}=0.6$
and $x=0$ and (d) $x=0.1$. Period doubling results in the appearance
of a subharmonic response even when $i_{4\pi}=0$, which may hide
the presence of a $4\pi\mbox{SC}$. It appears as a pitchfork bifurcation
at $i_{0}\simeq0.32$ in (c). The point at which this occurs has been
marked with a dashed magenta line. \label{fig:Bifurcation-diagram-for}}
\end{figure}
\begin{figure*}
\includegraphics[width=2\columnwidth]{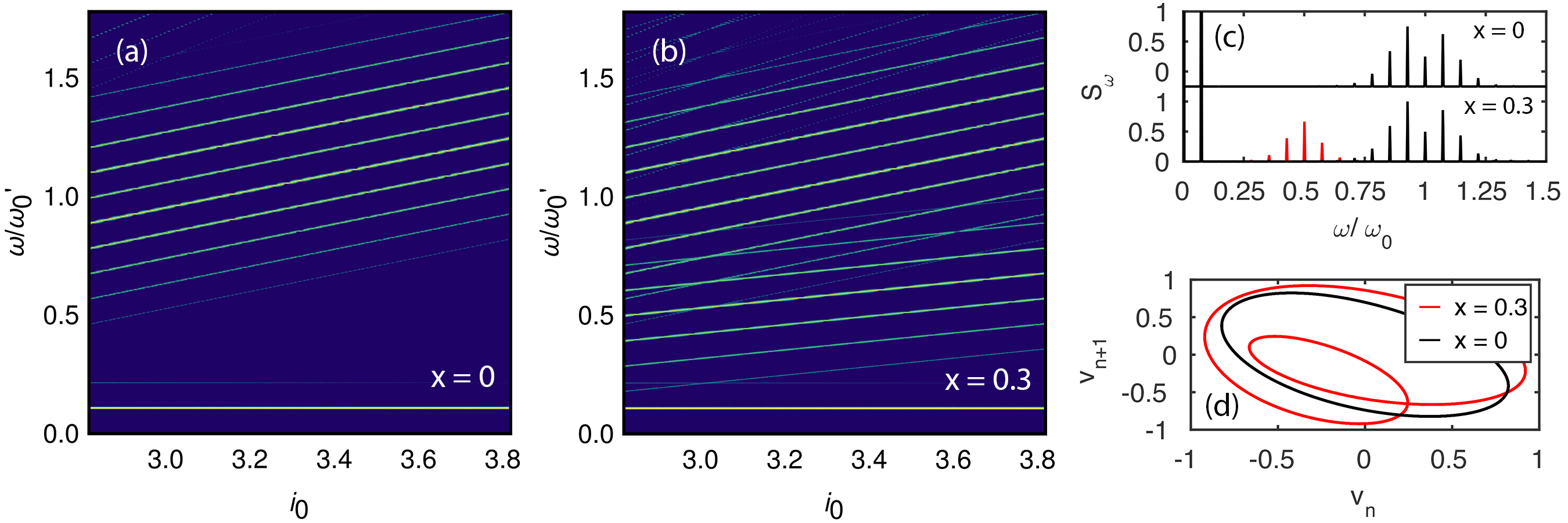}

\caption{(a), (b) Density plot of $\mbox{ln}[S_{\omega}(i_{0})]$ for $\omega_{\mathrm{ac}}=0.3$
and $i_{1}=0.58$, with a $4\pi$-periodic contribution of $x=0$
and $x=0.3$, respectively. In both cases, the frequency axis is normalized
to the natural frequency at the lowest value of $i_{0}$; that is,
$\omega_{0}'=\omega_{0}(2.82)$ (c) Fourier spectrum for the same
parameters as in (a) without a $4\pi\mbox{SC}$ (up) and with a $4\pi\mbox{SC}$
of $x=0.3$ (down), corresponding to $i_{0}=3.62$. The peaks corresponding
to the natural frequency and half of the natural frequency have been
indicated; the peaks originating from the $4\pi$-periodic contribution
are marked in red. (d) First return map (FRM) for the same parameters
as in (c) with black indicating the FRM for $x=0$ and red indicating
the FRM for $x=0.3$. \label{fig:Poincar=0000E9-maps-for-1}}
\end{figure*}

\subsection{Shapiro steps: period doubling}

The typical voltage response in the phase-locked regime of a topologically
trivial Josephson junction exhibits peaks at integer values of the
ac bias frequency, that is, at $\omega=k\omega_{\mathrm{ac}}$, $k\mathcal{2}\mathbb{Z}$.
However, for a finite $4\pi\mbox{SC}$, the voltage response inside
a given odd step is necessarily $4\pi$-periodic, as shown in Appendix~\ref{sec:Subharmonic-response-in},
while the even steps remain unaltered. This is shown, for example,
in the insets of Fig.~\ref{fig:remnant}. The time evolution of the
voltage on the $n$th \emph{odd }Shapiro step can be expressed as
a Fourier series
\[
v(t)=\sum_{l=0}^{\infty}v_{l}\mbox{e}^{i(l/2)\omega_{\mathrm{ac}}t},
\]
where $v_{0}=n\omega_{\mathrm{ac}}$. Then, the periodicity of the
voltage can be extracted from the emission spectrum. However, in a
capacitively-shunted junction, the non-linear dynamics can break the
symmetry of the RCSJ equation of motion. Then, even in absence of
$4\pi\mbox{SC}$, i.e: $x=0$, the steps may develop spontaneously
a subharmonic response at half the ac bias frequency, and thus the
Fourier spectrum shows peaks at integer multiples of $\omega_{\mathrm{ac}}/2$.
This phenomenon is called period doubling. It has been studied in
the past~\citep{Kautz1985,Kautz1987} in the context of the onset
of chaos in the RCSJ model~\citep{Pedersen1980}.

In Fig.~\ref{fig:Bifurcation-diagram-for} (a) and (b) we illustrate
the Fourier spectrum $v(\omega)$ as a function of $i_{0}$ for $x=0$
and $x\neq0$. In the interval $i_{0}\mathcal{2}[0.15,0.32]$, we
observe a set of resonances placed at integer multiples of the frequency
$\omega_{\mathrm{ac}}$ ($\omega_{\mathrm{ac}}/2)$ for $x=0$ ($x\neq0$).
In contrast, for $i_{0}>0.32$, the system becomes unstable and the
response becomes $4\pi$-periodic even for $x=0$, as the system enters
in a resonance at half the plasma frequency of the junction~\citep{Kautz1985}. 

The period doubling of a phase-locked solution manifests itself as
a pitchfork bifurcation inside the steps, whereas for a junction with
$i_{4\pi}\neq0$ the whole step is $4\pi$-periodic. This is represented
in Fig.~\ref{fig:Bifurcation-diagram-for} (c) and (d), where we
have plotted for each value of $i_{0}$ the voltage at different periods;
that is, for each $i_{0}$, we have plotted the set $\{v(t_{i})\}$,
where $t_{i+1}=t_{i}+2\pi/\omega_{\mathrm{ac}}$. For a $2\pi$-periodic
signal, to each $i_{0}$ corresponds a single value $v(t_{i})$, whereas
for a $4\pi$-periodic signal, to each $i_{0}$ corresponds two values
$\{v(t_{i}),v(t_{i}+2\pi/\omega_{\mathrm{ac}})\}$. The dense black
portions of the diagram correspond to aperiodic solutions. For $x\neq0$,
the whole step is $4\pi$-periodic, whereas for $x=0$, the step is
$4\pi$-periodic starting at the resonance at $i_{0}=0.32$. Then,
the experimental distinction between $x=0$ and $x\neq0$ requires
determining whether or not there is a peak in the Fourier spectrum
at $3\omega_{\mathrm{ac}}/2$ in the region $i_{0}\mathcal{2}[0.15,0.32]$. 

In Ref.~\citep{0034-4885-59-8-001}, the authors propose a condition
for avoiding period doubling, namely that the ac bias amplitude is
large enough that $\tilde{i}_{1}\gg\omega_{\mathrm{ac}}^{-4}$. Since
this corresponds to the Bessel regime, where we do not expect a large
$4\pi$-periodic response, \emph{the possibility of period doubling
cannot be neglected in the parameter regions where the junction is
strongly $4\pi$-periodic.}

\subsection{Quasiperiodic regime}

In order to explore other regimes, we may look instead at quasiperiodic
(QP) solutions appearing at $i_{0}\gg\omega_{\mathrm{ac}}\sigma\tilde{i}_{1}$~\citep{0034-4885-59-8-001}.
When the natural frequency of the junction $\omega_{0}=\sigma^{-1}i_{0}$
and the driving frequency $\omega_{\mathrm{ac}}$ are incommensurate,
the voltage response is quasiperiodic. This is the case in the apparently
continuous, linear parts of the $i_{0}-\mathcal{h}v\mathcal{i}$ curves,
as indicated in Fig.~\ref{fig:(A)-Scheme-of-1} (a), where the Shapiro
steps are small and the current follows the current-voltage characteristic.
The voltage response curve in the quasiperiodic regime of a Josephson
junction can be written as a generalized Fourier series~\citep{Schilder2006}
\[
v(t)=\overset{\infty}{\underset{k,l=0}{\sum}}v_{kl}e^{i(k\omega_{\mathrm{ac}}+l\omega_{0}/2)t},
\]
where $v_{kl}=\omega_{0}$. In the Fourier spectrum for $x=0$, shown
in Fig.~\ref{fig:Poincar=0000E9-maps-for-1} (a), there is a resonance
at $\omega_{\mathrm{ac}}$, corresponding to the ac bias and another
at $\omega_{0}$, corresponding to the intrinsic frequency of the
junction. Then, starting from $\omega_{0}$, another set of peaks
appears separated from $\omega_{0}$ an integer multiple of $\omega_{\mathrm{ac}}$.
The Fourier spectrum for the case $x\neq0$, represented in Fig.~\ref{fig:Poincar=0000E9-maps-for-1}
(a) exhibits an extra resonance at $\omega_{0}/2$ plus a new set
of satellite peaks once again separated an integer multiple of $\omega_{\mathrm{ac}}$.

The presence of a $4\pi\mbox{SC}$ can also be observed through the
first-return maps (FRMs), as these are heavily modified by change
in the periodicity of the supercurrent terms. A FRM is composed of
pairs $\{v(t_{i+1}),v(t_{i})\}$, where $t_{i+1}=t_{i}+2\pi/\omega_{\mathrm{ac}}$.
The FRMs are sensitive to the periodicity of the voltage response,
in a similar way to Poincaré maps~\citep{Luo1998}. The FRMs, however,
can be obtained from a time resolved scalar  response, like the voltage
signal of a typical Josephson junction experiment. For a $2\pi$-periodic
voltage response, the FRMs are ellipses, as can be seen in Fig.~\ref{fig:Poincar=0000E9-maps-for-1}
(d). As the periodicity of the response shifts to $4\pi$, however,
the ellipses twist inward and self-crossings appear, as represented
in Fig.~\ref{fig:Poincar=0000E9-maps-for-1} (d) for different values
of $i_{0}$. This is in accordance with the scenario observed in the
FRMs of superlattice current self-oscillations by Luo et al~\citep{Luo1998}. 

\emph{Both the changes to} \emph{the Fourier spectra and to the FRMs
of quasiperiodic solutions due to a finite $4\pi SC$ are a general
behavior when sufficiently away from the steps, and thus can be used
to discern the topological nature of the junction}. Close to the steps,
the solutions may be heavily distorted, exhibiting high subharmonic
response or even fractal structure in their FRMs~\citep{Bulashenko1999,Snchez2001}.

\section{Conclusions}

We have explained theoretically the experimental features observed
in a Josephson junction in the presence of a $4\pi\mbox{SC}$ by introducing
a capacitance term in the semiclassical equation of motion of the
junction. Namely, we have shown that in the underdamped regime, i.e:
when the capacitive term is stronger than the resistive one, the odd
steps are suppressed even for high ac bias amplitudes. Furthermore,
we observe an uneven quenching, with the first Shapiro step being
more affected by the presence of a $4\pi\mbox{SC}$ than the subsequent
odd steps. This behavior reproduces qualitatively the experimental
results published so far, and indicates that for a correct estimation
of the $4\pi\mbox{SC}$ amplitude it is necessary to consider the
presence of a finite capacitance in the junction. 

We also consider the possibility to study the periodicity of the junction
through the Fourier spectrum of the voltage response. We show how
the appearance of period doubling bifurcations in the regions where
the $4\pi$-periodic response is at its largest may difficult this
observation in the phase-locked regime, corresponding to the Shapiro
steps. The Fourier spectrum of quasiperiodic solutions also provides
information about the topology of the junction. While far from the
step regions, the quasiperiodic response is surprisingly stable and
shows the marks of a finite $4\pi$-periodic response in its Fourier
components. The corresponding first-return maps are twisted, compared
to the ellipses found for the $2\pi$-periodic case. This may be used
to discern the periodicity of the junction directly from the voltage
response.

Overall, we have analyzed both the phase-locked and the quasiperiodic
regimes of the topological RCSJ model, showing how each may be used
to help in the detection of Majorana bound states. Furthermore, the
results shown here for the RCSJ model may be useful to understand
the process of periodicity change in other non-linear systems.
\begin{acknowledgments}
This work was supported by the Spanish Ministry of Economy and Competitiveness
via Grant No. MAT2014- 58241-P and the Youth European Initiative together
with the Community of Madrid, Exp. PEJ15/IND/AI-0444. F. D. acknowledges
financial support from the DFG via SFB 1170 \textquotedbl{}ToCoTronics\textquotedbl{},
the Land of Bavaria (Institute for Topological Insulators and the
Elitenetzwerk Bayern).
\end{acknowledgments}

\appendix

\section{Subharmonic response in the $4\pi$-periodic Josephson junction.
\label{sec:Subharmonic-response-in}}

Phase lock occurs when the phase advances by $2\pi l$ after $m$
periods $T=2\pi/\omega_{\mathrm{ac}}$ of the ac driving
\begin{equation}
\varphi\left(t+mT\right)=\varphi(t)+2\pi l\label{eq:PL condition}
\end{equation}
for integers $l,m$. The Josephson equations predicts for that case
$\mathcal{h}v\mathcal{i}=\left\langle \mbox{d}\varphi/\mbox{d}t\right\rangle =(l/m)\omega_{\mathrm{ac}}$.
When $m>1$ and $l/m$ is not an integer, the junction may develop
subharmonic phase-lock, observed as steps corresponding to fractions
of the ac bias frequency. Subharmonic phase-locking is forbidden in
the RSJ limit~\citep{Waldram1982}. 

If the odd steps do not vanish completely, the voltage response inside
a given odd step will have twice the period of the ac bias, that is,
it will have $m=2$ in Eq.~(\ref{eq:PL condition}). The voltage
response corresponding to the first step may be $4\pi$-periodic and
still contribute to the step, provided that $\varphi\left(t+2T\right)=\varphi(t)+4\pi$.
In fact, as we will show now, the response inside a given odd step
\emph{must be $4\pi$-periodic}.

Consider a $2\pi$-periodic trial solution for the $n$th step of
the form

\begin{equation}
\varphi(t)=\varphi_{0}+n\omega_{\mathrm{ac}}t-\sum_{l=1}^{\infty}\tilde{i}_{l}\mathrm{sin}\left(l\omega_{\mathrm{ac}}t+\theta_{l}\right),\label{eq:trialfunc}
\end{equation}
with $n\mathcal{2\mathbb{Z}}$. Substituting Eq.~(\ref{eq:trialfunc})
into the RCSJ equation one obtains a complicated formula for the free
parameters $\varphi_{0},\{\tilde{i}_{l},\theta_{l}\}$. This can be
solved in particular limits, to yield analytical expressions for these
parameters, as detailed below. Here, instead, we integrate all terms
in the RCSJ equation over one period of the ac bias, obtaining 
\begin{equation}
\begin{array}{c}
i_{4\pi}\sum_{\mathbf{k}}\frac{2}{\left(\sum_{l}lk_{l}-\frac{n}{2}\right)\omega_{\mathrm{ac}}}\left[\prod_{l}J_{k_{l}}\left(\frac{\tilde{i}_{l}}{2}\right)\right]\\
\times\mathrm{sin}\left[\left(\sum_{l}lk_{l}-\frac{n}{2}\right)\pi+\frac{\varphi_{0}}{2}-\sum_{l}k_{l}\theta_{l}\right]\\
\times\mathrm{sin}\left[\left(\sum_{l}lk_{l}-\frac{n}{2}\right)\pi\right]=0,
\end{array}\label{eq:impulse}
\end{equation}
where $J_{n}(x)$ is the $n$th order Bessel function and $\mathbf{k}=(k_{1},k_{2},\ldots)$
is a vector of indices, each running from $-\infty$ to $\infty$.
Note that the sine factor multiplying the left-hand side is
\[
\mathrm{sin}\left[\left(\sum_{l}lk_{l}-\frac{n}{2}\right)\pi\right]=\left\{ \begin{array}{c}
0\,\,\,\mbox{if }n\mbox{ is even}\\
1\,\,\,\mbox{if }n\mbox{ is odd}
\end{array}\right.
\]

Thus, Eq.~(\ref{eq:impulse}) indicates that in the odd steps the
RCSJ equation cannot be satisfied if the solution is $2\pi$-periodic
even on average over a period. In order to obtain a satisfactory solution
for the RCSJ equation on an odd step, we need to take into account
the possibility of a subharmonic response, and include terms at frequencies
$l\omega_{\mathrm{ac}}/2$, $l\mathcal{2}\mathbb{Z}$.

\begin{figure}
\begin{centering}
\includegraphics[width=1\columnwidth]{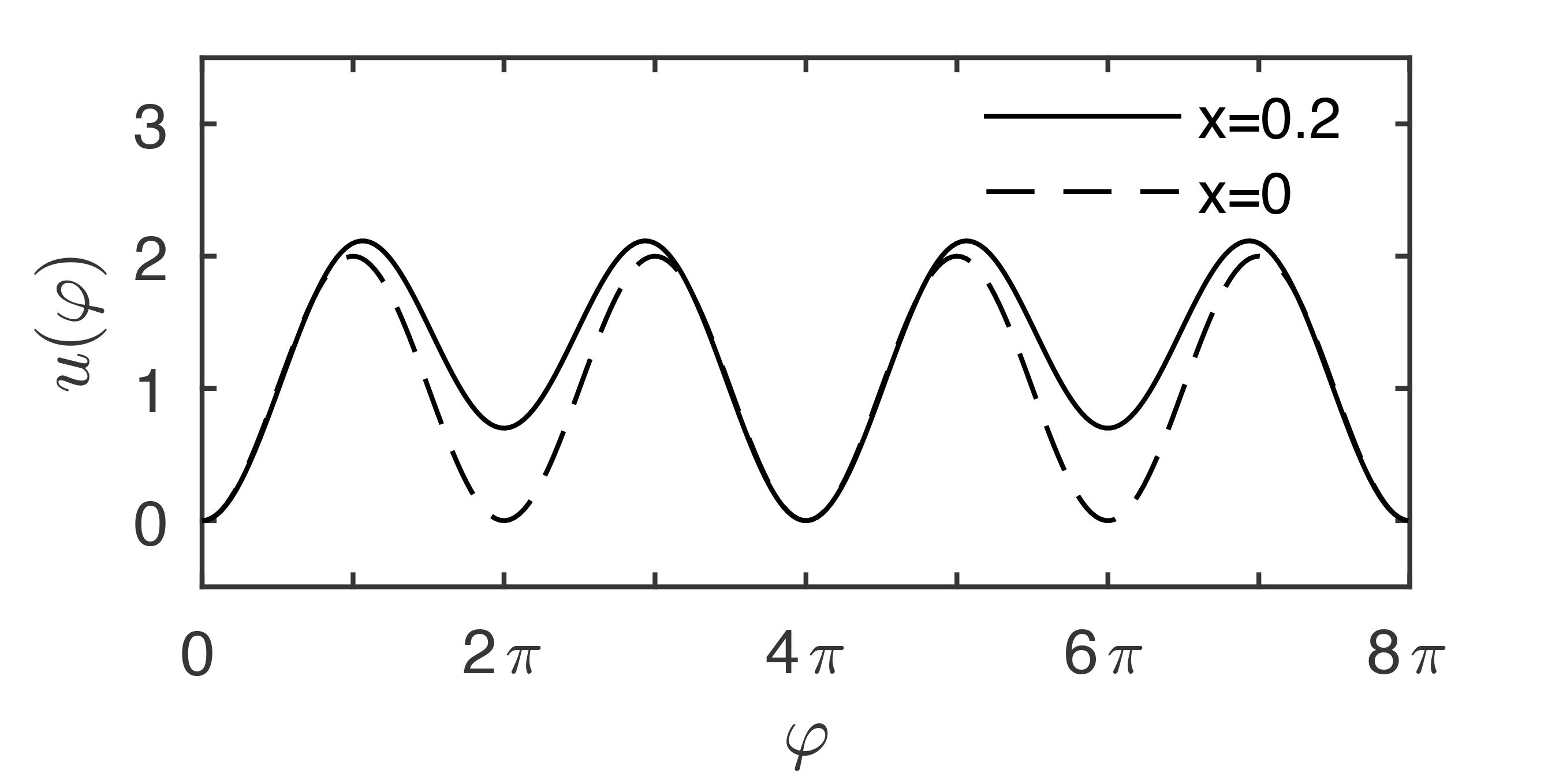}
\par\end{centering}

\caption{Washboard potential for $x=0$ (dashed lines) and $x=0.2$ (full lines).
The shape of the potential shifts its periodicity as the $4\pi$-periodic
contribution increases in importance. The result is two subsets of
wells: one consisting of shallow wells and another of deep wells.
\label{fig:Washboard-potential-for}}
\end{figure}

This can be understood by using the mechanical analogue of the RCSJ
equation, that is, by considering the RCSJ equation as representing
the motion of a massive particle under a \emph{washboard potential}
\[
u(\varphi)=i_{2\pi}\mathrm{cos}(\varphi)+i_{4\pi}\mathrm{cos}(\varphi/2)
\]
with damping given by $\sigma\mbox{d}\varphi/\mbox{d}t$ and both
a constant force $i_{0}$ and a time-dependent force $i_{1}\mathrm{sin}(\omega_{\mathrm{ac}}t)$.
The washboard potential looks like a sequence of potential wells,
as represented in Fig.~\ref{fig:Washboard-potential-for}. For a
small value of $x$, there are two types of wells with different heights,
so that a shallow well is followed by a deep well and vice versa.
For an odd step, the ``particle'' traverses an odd number of wells
each ac bias period. Hence after an ac bias period it moves from a
shallow well into a deep well, and it takes it another period to move
from a deep well back into a shallow well. For a particle in an odd
step, the potential appears\emph{ }$4\pi$-periodic, since it takes
two periods to go back to the initial position. In the case of an
even step, where the particle traverses an even number of wells each
ac bias period, after a cycle it ends its movement in the same type
of well it started. For a particle in an even step, the potential
\emph{appears} $2\pi$-periodic. The voltage response in an odd step,
will develop a $4\pi$-periodic voltage response, as it requires two
ac bias periods for the particle to turn back to the well where it
started. In that sense, the remainder in Eq.~(\ref{eq:impulse})
acts like an \emph{impulse} per ac bias period on the particle which
will tend to make its movement $4\pi$-periodic if the particle is
in an odd step, and will have no effect if the particle is in an even
step.

\section{Phase-lock in the Bessel regime\label{sec:Phase-lock-in-the}}

In this section, we obtain analytical expressions for the step widths
inside the Bessel regime. In that case, the voltage response to the
applied current bias may be approximated as linear $v(t)=v_{0}-v_{1}\mathrm{cos}(\omega_{\mathrm{ac}}t)$,
with $n\mathcal{2\mathbb{Z}}$. In that case, one obtains a trial
phase-locked solution $\varphi(t)=\varphi_{0}+v_{0}t-\tilde{i}_{1}\mathrm{sin}(\omega_{\mathrm{ac}}t+\theta_{1})$,
where $\tilde{i}_{1}=v_{1}/\omega_{\mathrm{ac}}$. Note that, as explained
above, this solutions cannot be exact for an odd step, and a $4\pi$-periodic
term needs to be added. We will consider this problem in Appendix~\ref{sub:The-high-ac}. 

Starting from $\varphi(t)$, one can calculate the amplitude of the\emph{
}steps in the Bessel regime~\citep{0034-4885-59-8-001} by substituting
$\varphi(t)$ and solving for $\varphi_{0}$. To do so, we equate
the \emph{constant} terms at both sides of the RCSJ equation with
$\varphi(t)$ substituted on it. Substitution inside the supercurrent
terms gives, according to the Jacobi-Anger expansion
\begin{equation}
\begin{array}{c}
i_{\mathrm{sc}}(t)=i_{2\pi}\sum_{l=-\infty}^{\infty}J_{l}(\tilde{i}_{1})\mathrm{sin}[(v_{0}-l\omega_{\mathrm{ac}})t+\varphi_{0}-l\theta_{1}]\\
+i_{4\pi}\sum_{l=-\infty}^{\infty}J_{l}(\tilde{i}_{1}/2)\mathrm{sin}[(v_{0}/2-l\omega_{\mathrm{ac}})t+\varphi_{0}/2-l\theta_{1}].
\end{array}\label{eq:supercurrentBessel}
\end{equation}

If $v_{0}\neq n\omega_{\mathrm{ac}}$, with $n$ an integer, the supercurrent
has no constant term. Equating the constant terms from the rest of
the RCSJ equation yields $v_{0}=i_{0}\sigma^{-1}$, indicating that
the voltage follows the resistive line $V=IR$. However, if $v_{0}=n\omega_{\mathrm{ac}}$
--that is, at the values that we expect Shapiro steps to appear--,
the supercurrent term\emph{ }contributes to the\emph{ average voltage.}
For a Shapiro step corresponding to $n$ odd, when $v_{0}=n\omega_{\mathrm{ac}}$
we obtain 

\begin{equation}
n\sigma\omega+i_{2\pi}J_{n}(\tilde{i}_{1})\mathrm{sin}\varphi_{n}=i_{0},\label{eq:step width odd bessel}
\end{equation}
where $\varphi_{n}=\varphi_{0}-n\theta_{1}$. This equation fixes
the free parameter $\varphi_{0}$. The interesting aspect of this
relation is that it is satisfied for a range of $i_{0}$ of 
\begin{equation}
(\Delta i_{0})_{n}^{\mathrm{odd}}=2i_{2\pi}|J_{n}(\tilde{i}_{1})|\label{eq:stepWidthOddBR}
\end{equation}
resulting in the appearance of a step at height $n\omega_{\mathrm{ac}}$,
as observed experimentally.

It remains to determine $\tilde{i}_{1}$ and $\theta_{1}$. To obtain
expressions for them one looks at the Fourier components at a frequency
of $\omega_{\mathrm{ac}}$. Ignoring the contribution from the terms
$\mathrm{sin}(\varphi)$ and $\mathrm{sin}(\varphi/2)$, one obtains
approximate expression for $\tilde{i}_{1}$ and $\theta_{1}$
\begin{equation}
\tilde{i}_{1}=\frac{i_{1}}{\omega_{\mathrm{ac}}\sqrt{\omega_{\mathrm{ac}}^{2}+\sigma^{2}}},\label{eq:i1tilde}
\end{equation}
\begin{equation}
\theta_{1}=\mathrm{arctan}(\sigma/\omega_{\mathrm{ac}}).\label{eq:theta1}
\end{equation}

As explained below, the approximation of neglecting the supercurrent
is justified provided that we stay in the Bessel regime. Even outside,
these definitions provide a satisfactory descriptions of the dependance
of $\tilde{i}_{1}$ and $\theta_{1}$ on $\omega_{\mathrm{ac}}$ and
$\sigma$. 

For the even steps, the analogue of Eq. (\ref{eq:step width odd bessel})
is
\begin{equation}
\begin{array}{c}
n\omega\sigma+i_{2\pi}J_{n}(\tilde{i}_{1})\mathrm{sin}\varphi_{n}\\
+i_{4\pi}J_{\frac{n}{2}}\left(\frac{\tilde{i}_{1}}{2}\right)\mathrm{sin}\left(\frac{1}{2}\varphi_{n}\right)=i_{0}
\end{array}\label{eq:step width even bessel}
\end{equation}
and the step width of the even steps is

\begin{equation}
\begin{array}{c}
(\Delta i_{0})_{n}^{\mathrm{even}}=2\mathrm{max}_{\varphi_{n}}\left[i_{2\pi}J_{n}(\tilde{i}_{1})\mathrm{sin}\varphi_{n}+\right.\\
\left.i_{4\pi}J_{\frac{n}{2}}\left(\frac{\tilde{i}_{1}}{2}\right)\mathrm{sin}\left(\frac{1}{2}\varphi_{n}\right)\right].
\end{array}\label{eq:stepWidthEvenBR}
\end{equation}

Except for the renormalization of $\tilde{i}_{1}$, these results
are consistent with those obtained in Ref.~\citep{1701.07389} within
the RSJ model, suggesting that the effect of capacitance is not so
important within the Bessel regime. Eq. ~\ref{eq:stepWidthOddBR}
indicates that the odd steps disappear at $i_{2\pi}=0$, while, according
to Eq.~\ref{eq:stepWidthEvenBR}, the even steps do not vanish as
the contribution coming from the $\propto i_{4\pi}$ term compensates
the decrease in the contribution from the $\propto i_{2\pi}$ term.
However, this trial solution neglects the fact that the voltage response
in an odd step has to be $4\pi$-periodic. In Appendix~\ref{sub:The-high-ac}
we will see that in the correct description (i.e: with a $4\pi$-periodic
voltage response) the odd steps have a finite width and hence do not
disappear completely.

\section{Outside the Bessel regime: step width\label{sec:Outside-the-Bessel-1}}

In this appendix we extend the results of the previous section to
parameter regions far from the Bessel regime. Here we cannot obtain
analytical results except for certain limits, but the trial function
method can be used to gain insight on the junction behavior. 

In order to study these effects, we study a quasiperiodic solution
of the type

\begin{equation}
\varphi(t)=\varphi_{0}+v_{0}t-\sum_{j,l=-\infty}^{\infty}\tilde{i}_{jl}\mathrm{sin}\left[\xi_{jl}t+\theta_{jl}\right],\label{eq:trial sol}
\end{equation}
where $\xi_{jl}=\left(jv_{0}/2+l\omega_{\mathrm{ac}}\right)$. We
also require that $\tilde{i}_{jl}=0$ if $\xi_{jl}=0$. Here, if $i_{4\pi}=0$
then $j$ is restricted to even numbers. This type of trial solution
replicates correctly the numerical results which show that the odd
steps develop a half harmonic response (i.e: the voltage response
is $4\pi$-periodic) whereas the even steps only exhibit integer harmonics.
We delay a proper justification for this trial function until the
end of this section. 

After inserting $\varphi(t)$ into the RCSJ equation, the supercurrent
is given by

\[
\begin{array}{c}
i_{\mbox{sc}}(t)=i_{2\pi}\sum_{\mathbf{k}}\mathcal{J}_{\mathbf{k}}(\tilde{\mathbf{i}})\mathrm{sin}\left[\varphi_{0}+\left(v_{0}-\mathbf{k}\cdot\boldsymbol{\xi}\right)t-\mathbf{k}\cdot\boldsymbol{\theta}\right]\\
+i_{4\pi}\sum_{\mathbf{k}}\mathcal{J}_{\mathbf{k}}(\tilde{\mathbf{i}}/2)\mathrm{sin}\left[\varphi_{0}/2+\left(v_{0}/2-\mathbf{k}\cdot\boldsymbol{\xi}\right)t-\mathbf{k}\cdot\boldsymbol{\theta}\right],
\end{array}
\]
where
\[
\mathcal{J}_{\mathbf{k}}(\tilde{\mathbf{i}})=\prod_{j,l}J_{k_{jl}}\left(\tilde{i}_{jl}\right)
\]
is a generalized Bessel function, $\mathbf{k}=(k_{ij})$ is a matrix
of indices and the sum is over all possible $\mathbf{k}$, with each
$k_{jl}$ going from $-\infty$ to $+\infty$. Similarly, $\tilde{\mathbf{i}}=(\tilde{i}_{ij})$
is a matrix of the Fourier amplitudes. We define in a similar way
$\boldsymbol{\xi}$ and $\boldsymbol{\theta}$. The dot indicates
the Frobenius inner product of matrices $\mathbf{a}\cdot\mathbf{b}=\sum_{jl}a_{jl}b_{jl}$. 

In a similar way to the Bessel regime, Shapiro steps appear around
certain values $v_{0}$ of the average voltage, satisfying

\begin{equation}
\begin{array}{c}
v_{0}-2\boldsymbol{\kappa}\cdot\boldsymbol{\xi}=0\\
v_{0}-\boldsymbol{\kappa}'\cdot\boldsymbol{\xi}=0
\end{array}\label{eq:averagevoltage1}
\end{equation}
for a set of indices $\{\kappa_{ij}\}$ and $\{\kappa_{ij}'\}$ running
from $-\infty$ to $+\infty$. The first (second) equation appears
as a result of the $2\pi$-periodic ($4\pi$-periodic) supercurrent
element. In terms of the ac bias frequency, steps appear at values
$v_{0}$ of the average voltage given by 
\begin{equation}
v_{0}=2\omega_{\mathrm{ac}}\frac{\sum_{j,l}\kappa_{jl}l}{2-\sum_{j,l}\kappa_{jl}j},\label{eq:stepsv0}
\end{equation}

\begin{equation}
v_{0}=2\omega_{\mathrm{ac}}\frac{\sum_{j,l}\kappa_{jl}'l}{1-\sum_{j,l}\kappa_{jl}'j}.\label{eq:stepsv0-1}
\end{equation}

Note that the $\{\kappa_{l,j}\}$ and $\{\kappa_{l,j}'\}$ coefficients
must be the same in both the denominator and the numerator. If $i_{2\pi}=0$,
then steps appear only at the values of $v_{0}$ given by Eq.~(\ref{eq:stepsv0-1}).
In that case, there is no reason to expect the odd steps to disappear
for a trial solution like Eq.~(\ref{eq:trial sol}). If $1-\sum_{j,l}\kappa_{jl}'j$
is even, then Eq.~(\ref{eq:stepsv0-1}) indicates that there will
be odd steps. This is the case represented in Fig.~\ref{fig:remnant}. 

The step width for a certain $v_{0}$ is obtained by finding the range
of $i_{0}$ for which there is a $\varphi_{0}$ that satisfies
\begin{equation}
\begin{array}{c}
\sigma v_{0}+i_{2\pi}\sum_{\boldsymbol{\kappa}}\mathcal{J}_{\boldsymbol{\kappa}}(\tilde{\mathbf{i}})\mathrm{sin}\left[\varphi_{0}-\boldsymbol{\kappa}\cdot\boldsymbol{\theta}\right]\\
+i_{4\pi}\sum_{\boldsymbol{\kappa}'}\mathcal{J}_{\boldsymbol{\kappa}'}(\tilde{\mathbf{i}}/2)\mathrm{sin}\left[\frac{\varphi_{0}}{2}-\boldsymbol{\kappa}'\cdot\boldsymbol{\theta}\right]=i_{0},
\end{array}\label{eq:stepwidth}
\end{equation}
with $\boldsymbol{\kappa}$ and $\boldsymbol{\kappa}'$ given by Eqs.~(\ref{eq:stepsv0})
and~\ref{eq:stepsv0-1}.

\section{Outside the Bessel regime: Fourier components\label{sec:Outside-the-Bessel}}

In this appendix we take the results of the previous section and focus
on the response inside a given odd step. For an even step, the trial
solution in Eq.~(\ref{eq:trial sol}) only has integer components.
For an odd step, the trial solution has integer and half-integer harmonics,
yielding

\[
\varphi(t)=\varphi_{0}+n\omega_{\mathrm{ac}}t-\sum_{l=1}^{\infty}\tilde{i}_{l/2}\mathrm{sin}\left(\frac{l\omega_{\mathrm{ac}}}{2}t+\theta_{l/2}\right).
\]

Then, the RCSJ equation becomes a set of equations for each pair of
Fourier amplitudes and phases $\{\tilde{i}_{l/2},\theta_{l/2}\}$.
Then, the previously defined $\mathbf{k},\tilde{\mathbf{i}},\boldsymbol{\theta},\boldsymbol{\xi}$
are vectors in the index $l$, so that, for example: $\mathbf{k}=(k_{1},k_{2},\ldots)$
We need these Fourier coefficients in order to obtain the step width
from Eq.~(\ref{eq:stepwidth}). In particular, for the Fourier component
at the frequency $m\omega_{\mathrm{ac}}/2$, there are two equations.

\begin{equation}
\begin{array}{c}
i_{2\pi}\sum_{\tilde{\mathbf{k}},s_{1}}s_{1}\mathcal{J}_{\tilde{\mathbf{k}}}(\tilde{\mathbf{i}})\mathrm{cos}\left(\varphi_{0}-\tilde{\mathbf{k}}\cdot\boldsymbol{\theta}\right)\\
+i_{4\pi}\sum_{\tilde{\mathbf{k}}',s_{1}}s_{1}\mathcal{J}_{\tilde{\mathbf{k}}'}(\tilde{\mathbf{i}}/2)\mathrm{cos}\left(\varphi_{0}/2-\tilde{\mathbf{k}}'\cdot\boldsymbol{\theta}\right)\\
+\frac{m}{2}\omega_{\mathrm{ac}}\sigma\tilde{i}_{m/2}\mathrm{sin}\left(\theta_{m/2}\right)+\frac{m^{2}}{4}\omega_{\mathrm{ac}}^{2}\tilde{i}_{m/2}\mathrm{cos}\left(\theta_{m/2}\right)=i_{1}\delta_{m,2},
\end{array}\label{eq:mth harmonic-1}
\end{equation}
\begin{equation}
\begin{array}{c}
i_{2\pi}\sum_{\tilde{\mathbf{k}},s_{1}}\mathcal{J}_{\tilde{\mathbf{k}}}(\tilde{\mathbf{i}})\mathrm{sin}\left(\varphi_{0}-\tilde{\mathbf{k}}\cdot\boldsymbol{\theta}\right)\\
+i_{4\pi}\sum_{\tilde{\mathbf{k}}',s_{1}}\mathcal{J}_{\tilde{\mathbf{k}}'}(\tilde{\mathbf{i}}/2)\times\mathrm{sin}\left(\varphi_{0}/2-\tilde{\mathbf{k}}'\cdot\boldsymbol{\theta}\right)\\
-\frac{m}{2}\omega_{\mathrm{ac}}\sigma\tilde{i}_{m/2}\mathrm{cos}\left(\theta_{m/2}\right)+\frac{m^{2}}{4}\omega_{\mathrm{ac}}^{2}\tilde{i}_{m/2}\mathrm{sin}\left(\theta_{m/2}\right)=0
\end{array}\label{eq:mth harmonic}
\end{equation}
with the term $i_{1}\delta_{m,2}$ coming from the ac bias. Here $s_{1}=\pm1$,
and the sum is over the values $\{\tilde{\mathbf{k}},\tilde{\mathbf{k}}'\}$
that satisfy
\begin{equation}
\begin{array}{c}
s_{1}m+\sum_{l}l\tilde{k}_{l}=2n,\\
s_{1}m+\sum_{l}l\tilde{k}_{l}'=n,
\end{array}\label{eq:condit1}
\end{equation}
corresponding to the terms proportional to $i_{2\pi}$ and to $i_{4\pi}$,
respectively.

If we follow the prescription given in the definition of $\tilde{i}_{1}$
and $\theta_{1}$ in Appendix~\ref{sec:Phase-lock-in-the}, we would
neglect the two sums coming from the supercurrent terms, and then
we would find the trivial solution $\tilde{i}_{m/2}=0$ for $m\neq2$
and we recover the expressions for $\tilde{i}_{1}$ and $\theta_{1}$
inside the Bessel regime, Eqs.~(\ref{eq:i1tilde}) and~(\ref{eq:theta1}).
Since this prescription is not valid whenever $\omega_{\mathrm{ac}}\sigma$
or $\omega_{\mathrm{ac}}^{2}$ are comparable to $i_{2\pi}$ or $i_{4\pi}$,
and $i_{2\pi}+i_{4\pi}\simeq1$, we obtain the previously stated result
that the system responds linearly to the applied bias whenever $\omega_{\mathrm{ac}}^{2}\gg1$
or $\omega_{\mathrm{ac}}\gg1/\sigma$. 

The $4\pi$-periodic supercurrent term has a strong effect on the
step widths when the terms it generates in Eqs.~(\ref{eq:mth harmonic-1})
and~(\ref{eq:mth harmonic}) are comparable to the rest of the terms.
That is
\[
\omega_{\mathrm{ac}}\sigma\lesssim i_{4\pi},\,\omega_{\mathrm{ac}}^{2}\lesssim i_{4\pi},
\]

At this point we can justify the choice of trial solution, Eq.~(\ref{eq:trial sol}),
and in particular the choice of the periodic part. Other trial solutions
are possible, such as transient solutions which may be of importance
for weak damping ($\sigma\ll1$). For that reason, the following reasoning
rests on the assumption that the non-periodic part of Eq.~(\ref{eq:trial sol})
is linear in time, resulting in phase-lock. 

It is clear that the Fourier components included in the trial solution
have to include a component at frequency $\omega_{\mathrm{ac}}$ for
any $i_{1}\neq0$. The first harmonic $\tilde{i}_{1}\mathrm{sin}(\omega_{\mathrm{ac}}t+\theta_{1})$,
together with the linear term $v_{0}t$ leads to supercurrent terms
of the form of Eq.~(\ref{eq:supercurrentBessel}). Then, consider
Eqs.~(\ref{eq:mth harmonic-1}) and~(\ref{eq:mth harmonic}) for
an arbitrary frequency $\omega_{x}$ and the related Fourier component
$\tilde{i}_{x}\mathrm{sin}(\omega_{x}t+\theta_{x})$. They show that
$\tilde{i}_{x}=0$ unless the supercurrent terms include a Fourier
component at frequency $\omega_{x}$. The supercurrent terms can be
written as a Fourier series with components at frequencies $v_{0}+l\omega_{\mathrm{ac}}$
and $v_{0}/2+l\omega_{\mathrm{ac}}$, $l\mathcal{2}\mathbb{Z}$, so
the only non-zero Fourier components appear at these frequencies.
Repeating this process with these new components --that is, taking
a trial solution with the Fourier components $v_{0}+l\omega_{\mathrm{ac}}$
and $v_{0}/2+l\omega_{\mathrm{ac}}$ and inserting it back into the
supercurrent terms-- leads to a Fourier series with components at
$jv_{0}/2+l\omega_{\mathrm{ac}}$, $j,l\mathcal{2}\mathbb{Z}$, which
again means that the only non-zero $\tilde{i}_{x}$ correspond to
these frequencies. Repeating this process once again gives no new
frequencies, justifying the terms retained in Eq.~(\ref{eq:trial sol}).

\section{The high ac bias amplitude limit \label{sub:The-high-ac}}

In this section we study the high ac-bias amplitude limit. We derive
conditions for the Bessel regime in terms of $i_{1}$. Then, we consider
an extension of the Bessel regime to accommodate a $4\pi$-periodic
response. We show that in that case the odd steps do not vanish completely. 

We assume that the $\tilde{i}_{l/2}$, $l\neq2$ are small. As a first
approximation, we may neglect all terms $\mathcal{O}(\tilde{i}_{l/2}\cdot\tilde{i}_{l'/2})$,
$l,l'\neq2$. For $x\ll1$, the Bessel function can be approximated
$J_{\alpha}(x)\sim(x/2)^{\alpha}\Gamma^{-1}(\alpha+1)$. Therefore,
this amounts to neglecting the Bessel functions of $\tilde{i}_{l/2},l\neq2$
at first order to obtain a self-consistent approximation in the sense
that it is valid only if the $\tilde{i}_{l/2},l\neq2$ obtained in
this way are small, up to an error of order $\mathcal{O}(\tilde{i}_{l/2}\cdot\tilde{i}_{l'/2})$,
$l,l'\neq2$. In this way Eqs.~(\ref{eq:mth harmonic-1}) and~(\ref{eq:mth harmonic})
become a linear system for the $\tilde{i}_{l/2},l\neq2$, with $\tilde{i}_{1}$
as a parameter determined directly by $i_{1}$. Then, the conditions
of Eq.~(\ref{eq:condit1}) are
\[
\begin{array}{c}
\tilde{k}_{2}=2n-s_{2}l-s_{1}m;\,\tilde{k}_{l}=1,\,\tilde{k}_{j}=0,\,j\neq2,l,\\
\tilde{k}_{2}'=n-s_{2}l-s_{1}m;\,\,\tilde{k}_{l}'=1,\,\tilde{k}_{j}'=0,\,j\neq2,l,\\
\tilde{k}_{1}=2n-s_{1}m,\,\tilde{k}_{j}=0,\,\forall j\neq1,\\
\tilde{k}_{1}'=n-s_{1}m,\,\tilde{k}_{j}'=0,\,\forall j\neq1,
\end{array}
\]
where $s_{2}=\pm1$, yielding, for the $m$th Fourier component
\begin{equation}
\begin{array}{c}
i_{2\pi}\sum_{l,s_{1},s_{2}}s_{1}\tilde{i}_{l/2}J_{\frac{2n-s_{2}l-s_{1}m}{2}}\left(\tilde{i}_{1}\right)\\
\times\mathrm{cos}\left[\varphi_{0}-l\theta_{l/2}-\frac{2n-s_{2}l-s_{1}m}{2}\theta_{1}\right]\\
+\frac{i_{4\pi}}{2}\sum_{l,s_{1},s_{2}}s_{1}\tilde{i}_{l/2}J_{\frac{n-s_{2}l-s_{1}m}{2}}\left(\frac{\tilde{i}_{1}}{2}\right)\\
\times\mathrm{cos}\left[\frac{1}{2}\varphi_{0}-l\theta_{l/2}-\frac{n-s_{2}l-s_{1}m}{2}\theta_{1}\right]\\
+m\omega_{\mathrm{ac}}\sigma\tilde{i}_{m/2}\mathrm{sin}\left(\theta_{m/2}\right)+\frac{m^{2}}{2}\omega_{\mathrm{ac}}^{2}\tilde{i}_{m/2}\mathrm{cos}\left(\theta_{m/2}\right)\\
=i_{2\pi}\sum_{s_{1}}s_{1}J_{\frac{2n-s_{1}m}{2}}\left(\tilde{i}_{1}\right)\mathrm{cos}\left[\varphi_{0}-\frac{2n-l-s_{1}m}{2}\theta_{1}\right]\\
+2i_{1}\delta_{m,2}+\frac{i_{4\pi}}{2}\sum_{s_{1}}s_{1}J_{\frac{n-s_{1}m}{2}}\left(\frac{\tilde{i}_{1}}{2}\right)\mathrm{cos}\left[\frac{\varphi_{0}}{2}-\frac{n-l-s_{1}m}{2}\theta_{1}\right],
\end{array}\label{eq:highacbias}
\end{equation}

\begin{equation}
\begin{array}{c}
i_{2\pi}\sum_{l,s_{1},s_{2}}\tilde{i}_{l/2}J_{\frac{2n-s_{2}l-s_{1}m}{2}}\left(\tilde{i}_{1}\right)\\
\times\mathrm{sin}\left[\varphi_{0}-l\theta_{l/2}-\frac{2n-s_{2}l-s_{1}m}{2}\theta_{1}\right]\\
+\frac{i_{4\pi}}{2}\sum_{l,s_{1},s_{2}}\tilde{i}_{l/2}J_{\frac{n-s_{2}l-s_{1}m}{2}}\left(\frac{\tilde{i}_{1}}{2}\right)\\
\times\mathrm{sin}\left[\frac{1}{2}\varphi_{0}-l\theta_{l/2}-\frac{n-s_{2}l-s_{1}m}{2}\theta_{1}\right]\\
-m\omega_{\mathrm{ac}}\sigma\tilde{i}_{m/2}\mathrm{cos}\left(\theta_{m/2}\right)+\frac{m^{2}}{2}\omega_{\mathrm{ac}}^{2}\tilde{i}_{m/2}\mathrm{sin}\left(\theta_{m/2}\right)\\
=i_{2\pi}\sum_{s_{1}}J_{\frac{2n-s_{1}m}{2}}\left(\tilde{i}_{1}\right)\mathrm{sin}\left[\varphi_{0}-\frac{2n-l-s_{1}m}{2}\theta_{1}\right]\\
+\frac{i_{4\pi}}{2}\sum_{s_{1}}J_{\frac{n-s_{1}m}{2}}\left(\frac{\tilde{i}_{1}}{2}\right)\mathrm{sin}\left[\frac{\varphi_{0}}{2}-\frac{n-l-s_{1}m}{2}\theta_{1}\right].
\end{array}\label{eq:highacbias-2}
\end{equation}

Note that the higher $m$ harmonics are less affected by the supercurrent
channels, having an effective bias frequency of $m\omega_{\mathrm{ac}}$.
Because the Bessel function of order $k$ decreases as $\tilde{i}_{1}^{-1/2}$
for $\tilde{i}_{1}\gg k$, the lower harmonics, other than $m=2$,
are suppressed at high ac bias. Then, the supercurrent terms may be
neglected and the linear voltage response approximation is again valid.
In particular, provided that 
\begin{equation}
\tilde{i}_{1}\gg\sigma^{-2}\omega_{\mathrm{ac}}^{-2},\,\tilde{i}_{1}\gg\omega_{\mathrm{ac}}^{-4}\label{eq:condi1}
\end{equation}
are satisfied, the Bessel regime is recovered and Eqs.~(\ref{eq:i1tilde})
and~(\ref{eq:theta1}) are valid. These results reproduce those obtained
by means of Lyapunov stability analysis in Ref.~\citep{0034-4885-59-8-001}.

As noted above, the Bessel regime is not a satisfactory description
of the linear response, as it assumes that the voltage response in
the odd steps is $2\pi$-periodic. This discrepancy can be solved
by including the next order contributions. In that regard, note that
the terms in the right hand side of Eqs.~(\ref{eq:highacbias}) and~(\ref{eq:highacbias-2})
are of higher order than the other neglected terms. That is, we may
consider that the conditions of Eq.~(\ref{eq:condi1}) are satisfied,
but 
\[
\tilde{i}_{m/2}^{2}\tilde{i}_{1}\cancel{\gg}\sigma^{-2}\omega_{\mathrm{ac}}^{-2},\,\tilde{i}_{m/2}^{2}\tilde{i}_{1}\cancel{\gg}\omega_{\mathrm{ac}}^{-4}.
\]

If the right hand side terms are kept as the next order approximation,
we see that the $\propto i_{2\pi}$ terms in the right hand side are
zero for $m$ odd while the $\propto i_{4\pi}$ terms are zero for
$m$ even. Thus, for $i_{2\pi}=0$, the only harmonics are half-harmonics,
coming from the $4\pi$-periodic contribution to the supercurrent
(apart, of course, from the $\tilde{i}_{1}$ term). Take $i_{2\pi}=0$.
Then, there are terms contributing to the odd $n$th step, which satisfy
\[
n-s_{1}l-2s_{2}\tilde{k}_{2}=0
\]
with $s_{i}=\pm1$, $i=1,2$. The next order contributions result
in an increase in the step width of the odd steps compared to the
purely $2\pi$-periodic result. This means that \emph{in the corrected
(i.e: with a $4\pi$-periodic response) Bessel regime at high ac bias
the odd steps do not completely vanish, even at $i_{2\pi}=0$}. The
step width of the odd steps is nonetheless small but not zero. This
is confirmed by numerical results, such as in Fig.~\ref{fig:remnant},
where both the first and third steps have a finite width at $i_{2\pi}=0$.
As represented in the insets of Fig.~\ref{fig:remnant}, the Fourier
spectrum at the first step consists of a peak at $\omega_{\mathrm{ac}}$
and components at odd multiples of $\omega_{\mathrm{ac}}/2$, whereas
the Fourier spectrum at the second step consists of integer multiples
of $\omega_{\mathrm{ac}}$, as obtained analytically. 

\bibliographystyle{revtex/bibtex/bst/revtex/apsrev4-1}
\bibliography{rcsj}

\end{document}